\documentclass[12pt,a4paper]{article}

\usepackage{graphicx}
\usepackage{epstopdf}
\usepackage{amscd}
\usepackage{amssymb}
\usepackage{amsmath}

\title{Adhesion of surfaces {\em via} particle adsorption: Exact results for a lattice of fluid columns}

\author{Bartosz R\'{o}\.{z}ycki, Reinhard Lipowsky and Thomas R. Weikl \\ 
Max Planck Institute of Colloids and Interfaces, \\ 
Department of Theory and Bio-Systems, 14424 Potsdam, Germany}

\begin{document}

\maketitle

\begin{abstract}
We present here exact results for a one-dimensional gas, or fluid, of hard-sphere particles with attractive boundaries. The particles, which can exchange with a bulk reservoir, mediate an interaction between the boundaries. A two-dimensional lattice of such one-dimensional gas `columns' represents a discrete approximation of a three-dimensional gas of particles between two surfaces. The effective particle-mediated interaction potential of the boundaries, or surfaces, is calculated from the grand-canonical partition function of the one-dimensional gas of particles, which is an extension of the well-studied Tonks gas \cite{tonks36}. The effective interaction potential exhibits two minima. The first minimum at boundary contact reflects depletion interactions, while the second minimum at separations close to the particle diameter results from a single adsorbed particle that crosslinks the two boundaries. The second minimum is the global minimum for sufficiently large binding energies of the particles. Interestingly, the effective adhesion energy corresponding to this minimum is maximal at intermediate concentrations of the particles.
\end{abstract}

\section{Introduction}

The interactions of surfaces are often affected by nanoparticles or macromolecules in the surrounding medium. Non-adhesive particles cause attractive depletion interactions between the surfaces, since the excluded volume of the molecules depends on the surface separation \cite{Asakura54, Dinsmore96, Anderson02}. Adhesive particles, on the other hand, can directly bind two surfaces together if the surface separation is close to the particle diameter \cite{Baksh04,Winter06,Hu04}. In a recent letter \cite{Rozycki08}, we have presented a general, statistical-mechanical model for two surfaces in contact with adhesive particles. In this model, the space between the surfaces is discretized into columns of the same diameter $d$ as the particles. The approximation implied by this discretization is valid for small bulk volume fractions of the particles, since three-dimensional packing effects relevant at larger volume fractions are neglected. For short-ranged particle-surface interactions, the gas of particles between the surfaces is as dilute as in the bulk for large surface separations, except for the single adsorption layers of particles at the surfaces.

In this article, we present an exact solution of the one-dimensional gas of hard-sphere particles in a single column between two `surfaces'. Our aim here is two-fold. First, the exact solution presented here corroborates our previous, approximate solution for this one-dimensional gas obtained from a virial expansion in the particle concentration \cite{Rozycki08}. Second, the exactly solvable, one-dimensional model considered here is a simple toy model to study the interplay of surface adhesion and particle adsorption. Exactly solvable, one-dimensional models have played an important role in statistical mechanics \cite{lieb66, baxter82}. One example is the Kac-Baker model \cite{Kac59, Beker61, Kac63}, which has shed light on the statistical origin of phase transitions of the classical van der Waals type. More recent examples are models for one-dimensional interfaces, or strings, which have revealed the relevance of entropy and steric interactions in membrane unbinding and wetting transitions \cite{lipo66, lipo120, rozycki03}. Other examples are the Tonks model \cite{tonks36} and its various generalizations \cite{gursey50, salsburg53, baur62, davis96}, which have influenced our understanding of the relations between short-ranged particle interactions, thermodynamics, and statistical correlations in simple fluids. The Tonks model has been exploited also in soft-matter physics to investigate structures of confined fluids \cite{davis90, henderson07}, depletion phenomena in two-component mixtures \cite{lekkerkerker00}, thermal properties of columnar liquid crystals \cite{wensink04} and the phase behavior of polydisperse wormlike micelles \cite{vanderSchoot96}. A recent biophysical modification of the Tonks model addresses the wrapping of DNA around histone proteins \cite{chou03}. The model considered here is a novel extension of the Tonks model. 

\begin{figure}[h]
\begin{center}
\resizebox{\columnwidth}{!}{\includegraphics{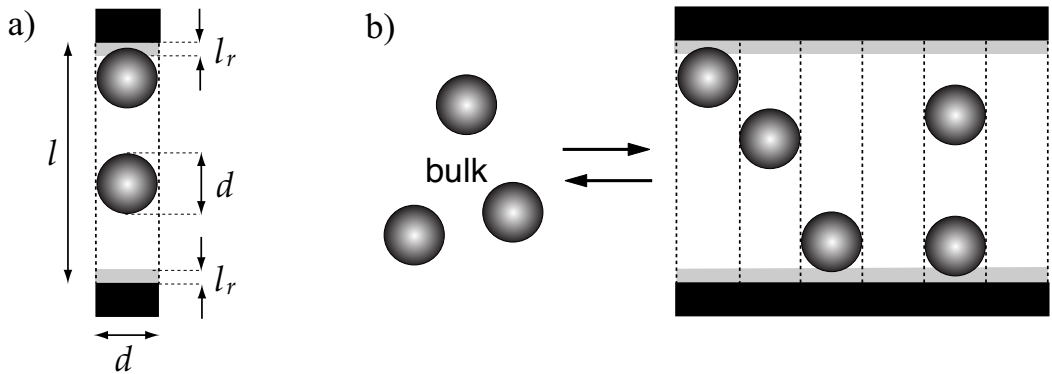}}
\caption{{\bf a)} A one-dimensional gas, or fluid, of hard-sphere particles in a column of the same diameter $d$ as the particles. The interaction between the particles and the boundaries, or `surfaces', is described by a square-well potential of depth $U$ and range $l_r<d/2$. A particle thus gains the binding energy $U$ if its center is  located at a distance smaller than $l_r +d/2$ from one of the surfaces. The length $\ell$ of the column corresponds to the separation of the surfaces. We consider the grand-canonical ensemble in which the particles in the column exchange with a bulk reservoir. -  {\bf b)} For small bulk volume fractions, a two-dimensional lattice of such columns represents a discrete approximation of a three-dimensional gas of particles between two surfaces \cite{Rozycki08}. Since the particle-surface interactions are short-ranged in our model, the particle gas between the surfaces is as dilute as in the bulk for large surfaces separations, except for the adsorption layer of particles at each of the surfaces.}
\label{cartoon}
\end{center}
\end{figure}

In our model, a one-dimensional gas of hard-sphere particles is attracted to the system boundaries, or `surfaces', by short-ranged interactions. We calculate the effective, particle-mediated interaction potential between the surfaces, $V$, by explicit integration over the particles' degrees of freedom in the partition function. The potential $V$ is a function of the surface separation $\ell$ and exhibits a minimum at surface contact, which reflects depletion interactions, and a second minimum at separations close to the diameter of the adhesive particles. The effective, particle-mediated adhesion energy of the surfaces, $W$, can be determined from the interaction potential $V$. The adhesion energy is the minimal work that has to be performed to bring the surfaces apart from the equilibrium state corresponding to the deepest well of the potential $V( \ell )$. Interestingly, the adhesion energy $W$ attains a maximum value at an optimal particle concentration in the bulk, and is considerably smaller both for lower and higher particle bulk concentrations. 

This article is organized as follows. In section~2, we introduce our model and define the thermodynamic quantities of interest. In section~3, we calculate the particle-mediated interaction potential $V( \ell )$ of the surfaces. The global minimum of this potential is determined in section~4, and the effective adhesion energy of the surfaces in section~5. In section~6, we show that the interaction potential $V( \ell )$ exhibits a barrier at surface separations slightly larger than the particle diameter, because a particle bound to one of the surfaces `blocks' the binding of a second particle to the apposing surface. The particle binding probability is calculated and analyzed in section~7. 

\section{Model and definitions \label{Definitions} }

We consider a one-dimensional gas of particles with attractive boundaries, see figure~\ref{cartoon}. The particles are modeled as hard spheres, and the attractive interaction between the particles and the boundaries, or `surfaces', is described by a square-well potential with depth $U$ and range $l_r$. The length $\ell$ of the gas `column' corresponds to the separation of the surfaces and the width of the column is chosen to be equal to the particle diameter $d$. The particles in the column exchange with a bulk reservoir of particles.

The position of the center of mass of particle $k$ is denoted by $x_k$, and its momentum by $p_k$. For the system of $n$ hard particles confined in the column of length $\ell > nd$, one has $d/2 < x_1, \quad x_1 < x_2 -d, \quad x_2 < x_3 - d, \quad \dots \quad x_n < \ell -d/2 $. We assume that the $1$-st and $n$-th particle interact with the surfaces, i.e.~with the bases of the columns, {\em via} the square-well potential
\begin{equation}
\mathcal{V}_{n} \left\{ x_k \right\} = - U \Theta \left( \frac{d}{2} +l_r - x_1 \right) - U \Theta \left( x_n - \ell +l_r +\frac{d}{2} \right) ,
\label{square-well-pot}
\end{equation}
where $U>0$ and $l_r>0$ are the potential depth and range, respectively. We also assume that $l_r < d/2$. Here and below, $\Theta$ denotes the Heaviside step function with $\Theta (x) =1$ for $x>0$ and $\Theta (x) =0$ for $x<0$. The configuration energy for the system of $n$ particles in the column is
\begin{equation}
\mathcal{H}_n \left\{ x_k, p_k \right\} =  \mathcal{V}_n \left\{ x_k \right\}  + \sum_{k=1}^{n} \frac{p_k^2}{2m}
\end{equation}
and the corresponding canonical partition function can be written as
\begin{equation}
\mathcal{Z}_n = \frac{1}{\Lambda^n} \int_{\left( n- \frac{1}{2} \right)d}^{\ell -\frac{d}{2}} {\rm d} x_n \int^{x_n -d}_{\left( n- \frac{3}{2} \right) d} {\rm d}x_{n-1} \dots \int^{x_3 -d}_{\frac{3}{2} d} {\rm d}x_2 \int_{\frac{1}{2} d}^{x_2 -d} {\rm d}x_1 e^{ - \mathcal{V}_n \left\{ x_k \right\} /T }
\label{cpf}
\end{equation}
after integration over the momenta of the particles, see, e.g., \cite{tonks36, davis90, chou03}. Here, $\Lambda =h/ (2 \pi m T)^{1/2}$ is the thermal de Broglie wavelength, and $T$ denotes the temperature times the Boltzmann constant. In other words, $T$ is the basic energy scale.

Since the particles can exchange with the bulk solution, the number $n$ of particles in the column is not constant. Such a system is described by the grand-canonical ensemble in which the temperature, the column length $\ell$, and the particle chemical potential $\mu$ are fixed. The corresponding grand-canonical partition function
\begin{equation}
\mathcal{Z} = 1+ \sum_{n=1}^{ \lfloor \ell /d \rfloor } \mathcal{Z}_n e^{n \mu /T}
\label{gcpf}
\end{equation}
is a sum of a finite number of elements, where $\lfloor \ell /d \rfloor$ denotes the largest integer less than or equal $\ell /d$. The upper limit of the sum on the right hand side of equation (\ref{gcpf}) is  the largest number of hard particles with diameter $d$ that can be accommodated in a column of length $\ell$. The partition function $\mathcal{Z}$ given by equation (\ref{gcpf}) determines the grand potential
\begin{equation}
F_{\rm gc} = - T \ln \mathcal{Z} ,
\label{fe}
\end{equation}
the bulk density of the grand potential
\begin{equation}
f_{\rm gc} = \lim_{\ell \to \infty} \frac{F_{\rm gc} \, d}{\ell}
\label{bfed}
\end{equation}
and, hence, the surface contribution to the grand potential $F_{\rm gc}^{\rm (s)} = F_{\rm gc} - f_{\rm gc} \, \ell /d$. The effective interaction potential of the surfaces~\footnote{The surface interaction potential $V$ was called the `effective adhesion potential' in reference~\cite{Rozycki08}.},
\begin{equation}
V  = \frac{F_{\rm gc}^{\rm (s)}}{d^2} = \frac{ F_{\rm gc} \, d - f_{\rm gc} \, \ell }{d^3} ,
\label{sfe}
\end{equation}
is defined as the density of the surface contribution to the grand potential $F_{\rm gc}$. For consistency with our previous model for particle-mediated surface interactions \cite{Rozycki08}, the column bases are chosen here to be squares of side length $d$. Thus the $d^2$ in the denominator of equation~(\ref{sfe}) is the column base area. The surface potential $V$ defined by equation~(\ref{sfe}) is the main quantity of interest here and will be determined in the next section.

\section{Effective surface interaction potential}

Equations (\ref{gcpf}) - (\ref{sfe}) imply that
\begin{equation}
\exp \left( - V \frac{d^2}{T} \right) = \left( 1+ \sum_{n=1}^{ \lfloor \ell /d \rfloor } \mathcal{Z}_n e^{n \, \mu /T} \right) \exp \left( \frac{f_{\rm gc} \, \ell}{T \, d} \right) .
\label{V}
\end{equation}
To determine the surface interaction potential $V$, we thus have to calculate the canonical partition function $\mathcal{Z}_n$ and the bulk density of the grand potential, $f_{\rm gc}$, defined in equation (\ref{bfed}). The $n$-particle partition function $\mathcal{Z}_n$ is defined in equation (\ref{cpf}). The change of variables $y_k = x_k - \left( k- \frac{1}{2} \right) d$ in equation~(\ref{cpf}), with $k=1,2 \dots ,n$, leads to
\begin{equation}
\mathcal{Z}_n = \frac{1}{\Lambda^n} \int_{0}^{\ell -nd} {\rm d} y_n e^{U \Theta \left( y_n - \ell + nd + l_r \right) /T} \int_{0}^{y_n} {\rm d}y_{n-1} \dots \int_{0}^{y_3} {\rm d}y_2 \int_{0}^{y_2} {\rm d} y_1 e^{U \Theta \left( l_r - y_1 \right) /T} ,
\label{integral_Zn}
\end{equation}
where $U$ is the binding energy of the particles. The integral (\ref{integral_Zn}) can be evaluated, see \ref{appendix_cpf}, and after some computation we arrive at
\begin{equation}
\mathcal{Z}_n =  \left( e^{U/T} - 1 \right)^2 \, \Phi_n \left( \ell -2 l_r \right) - 2 e^{U/T} \left( e^{U/T} -1\right) \, \Phi_n \left( \ell - l_r \right) + e^{2U/T} \, \Phi_n \left( \ell \right) ,
\label{Zn}
\end{equation}
with
\begin{equation}
\Phi_n \left( l_0 \right) = \frac{1}{n!} \left( \frac{l_0 - nd}{ \Lambda } \right)^n \, \Theta \left( l_0 - nd \right)
\label{Phi_n}
\end{equation}
for any length $l_0$.  For $U=0$, equation (\ref{Zn}), reduces to the partition function
\begin{equation}
\mathcal{Z}_n = \frac{1}{\Lambda^n n!} \left( \ell - nd \right)^n \, \Theta \left( \ell - nd \right)
\end{equation}
of the classical Tonks gas \cite{tonks36}. The grand potential density $f_{\rm gc}$ can be derived from this exact result as shown in the following subsection.

\subsection{Thermodynamic potentials in the bulk}

The canonical and grand-canonical ensembles are equivalent in the thermodynamic limit, i.e.~for infinite surface separation $\ell$. In this limit, the grand potential density $f_{\rm gc}$ therefore can be obtained from the canonical potential density {\em via} Legendre transformation. First, we define the canonical free energy $F_{\rm ca} = -T \ln \mathcal{Z}_n$ and the free energy density in the bulk,
\begin{equation}
f_{\rm ca}  = \lim_{\infty} \frac{F_{\rm ca} \, d}{\ell} ,
\label{fca_definition}
\end{equation}
where $\lim_{\infty}$ denotes the thermodynamic limit in which both the column length $\ell$ and the particle number $n$ go to infinity while the particle `volume fraction'
\begin{equation}
\phi = \frac{n \, d}{\ell}
\label{phi_definition}
\end{equation}
remains constant. The particle volume fraction in one dimension defined by equation~(\ref{phi_definition}) attains values $0 \le \phi \le 1$, and could also be called a `length fraction'. In the one-dimensional model considered here, close packing corresponds to $\phi = 1$. 

The free energy density $f_{\rm ca}$ defined by equation~(\ref{fca_definition}) is an intensive quantity in the thermodynamic limit and, therefore, does not depend on the boundary conditions. In particular, the free energy density $f_{\rm ca}$ and its derivatives do not depend on the boundary potential (\ref{square-well-pot}), which is characterized by the binding energy $U$ and range $l_r$.
With the exact expression for the canonical partition function $\mathcal{Z}_n$ given in equation (\ref{Zn}) and the Sterling formula
\begin{equation}
\ln (n!) \approx n \ln n - n + \frac{1}{2} \ln \left( 2 \pi n \right) ,
\label{Stirling}
\end{equation}
we get
\begin{equation}
f_{\rm ca} = T \phi \left[ -1 + \ln \left( \frac{\Lambda}{d} \frac{\phi}{1 - \phi} \right) \right] .
\label{ffb}
\end{equation}
As expected, the particle-surface interactions characterized by the binding energy $U$ and range $l_r$ do not affect the free energy density $f_{\rm ca}$ in the thermodynamic limit. 
From the canonical free energy density $f_{\rm ca}$, we obtain the chemical potential
\begin{equation}
\mu = \left( \frac{\partial f_{\rm ca}}{ \partial \phi} \right)_{T} 
\label{mu}
\end{equation}
for the particles in the bulk. Equations (\ref{ffb}) and (\ref{mu}) lead to
\begin{equation}
\mu = T \ln \left( \frac{\Lambda}{d} \frac{\phi}{1 -\phi} \right) + \frac{T \, \phi}{1 -\phi}
\label{muu}
\end{equation}
which can be rewritten as
\begin{equation}
e^{\mu /T} = \frac{\Lambda}{d} \frac{\phi}{1 -\phi} \exp \left( \frac{\phi}{1 - \phi } \right) .
\label{muuu}
\end{equation}
Finally, the grand potential density $f_{\rm gc}$ follows from the Legendre transformation
\begin{equation}
f_{\rm gc} = f_{\rm ca} - \mu \, \phi .
\label{Legendre}
\end{equation}
Equations (\ref{ffb}), (\ref{muu}) and (\ref{Legendre}) lead to
\begin{equation}
f_{\rm gc} = - \frac{ T \phi}{1- \phi} ,
\label{omega}
\end{equation}
with the dependence between the chemical potential $\mu$ and the particle bulk volume fraction $\phi$ given in equation~(\ref{muu}). Basic thermodynamics implies that the gas pressure in the bulk is $p= - f_{\rm gc} /d^3$. This leads to the pressure $p = T \phi /[ (1- \phi ) d^3]$ here, which is the correct equation of state for the Tonks gas \cite{tonks36}.

\subsection{Potential profile}

\begin{figure}[h]
\begin{center}
\resizebox{0.7\columnwidth}{!}{\includegraphics{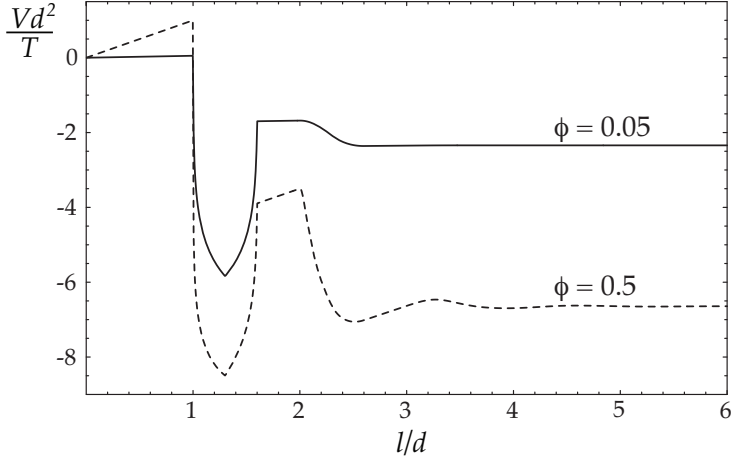}}
\caption{Rescaled surface interaction potential $V d^2 /T$, given by equation (\ref{VV}), as a function of the rescaled surface separation $\ell /d$ where $d$ is the particle diameter and $T$ denotes the temperature in energy units. The particle binding energy here is $U=5 \,T$, the binding range is $l_r =0.3 \, d$, and the particle bulk volume fraction is $\phi =0.05$ (solid line) and $\phi =0.5$ (dashed line), respectively. For large surface separations $\ell \gg d$, the potential $V( \ell )$ attains an approximately constant value $V_{\infty}$. According to equation (\ref{Vinfty}), the asymptotic values are $V_{\infty} \approx -6.6446 \, T/d^2$ for $\phi =0.5$ (dashed line) and $V_{\infty} \approx -2.3422 \, T/d^2$ for $\phi =0.05$ (solid line), in excellent agreement with numerical values obtained from equation (\ref{VV}). }
\label{V1}
\end{center}
\end{figure}

By combining equations (\ref{V}), (\ref{muuu}) and (\ref{omega}), the effective, particle-mediated interaction potential $V$ of the surfaces can be expressed as a function of the particle bulk volume fraction $\phi$ and the separation $\ell$ between the surfaces:
\begin{equation}
V = - \frac{T}{d^2} \ln \left[ 1+ \sum_{n=1}^{ \lfloor \ell /d \rfloor } \mathcal{Z}_n  \left( \frac{\Lambda}{d} \frac{\phi}{1 -\phi} \right)^n \exp \left( \frac{n \, \phi}{1 - \phi } \right) \right] + \frac{ \ell \, T}{d^3} \frac{\phi}{1 - \phi} ,
\label{VV}
\end{equation}
With the exact expression for the canonical partition function $\mathcal{Z}_n$ given in equations (\ref{Zn}) and (\ref{Phi_n}), the potential $V$ can be evaluated numerically for any finite surface separation $\ell$, see figures~\ref{V1} and \ref{V2}. For large bulk volume fractions $\phi$ (dashed curve in figure~\ref{V1}), the potential $V$ exhibits oscillations up to surface separations $\ell$ of the order of several particle diameters before approaching a constant, asymptotic value $V_{\infty}$ for large surface separations. The oscillations are related to successive layers of particles formed in the space between the surfaces. For small bulk volume fractions $\phi$, in contrast, the interaction potential $V$ attains an approximately constant value for surface separations $\ell > 2(d+ l_r)$, see solid curve in figure~\ref{V1}. 
%

\subsection{Potential asymptote}

Since the interactions between the particles and the surfaces are short-ranged, the potential $V$ has a horizontal asymptote, i.e.~$V(\ell)$ approaches a constant value $V_{\infty}$ for large surface separation $\ell$, see figures~\ref{V1} and \ref{V2}. In this subsection, we calculate the position of the  asymptote. To simplify the notation, we first introduce the auxiliary variable
\begin{equation}
\zeta = \frac{\phi}{1- \phi} ,
\label{zeta}
\end{equation}
%
and the function
\begin{equation}
{\rm g} \left( \frac{\ell}{d} , \frac{l_0}{d} , \zeta \right)= \sum_{n=1}^{\lfloor \ell /d \rfloor} \frac{\zeta^n}{n!} \exp \left( - \frac{\ell - nd}{d} \, \zeta \right) \left( \frac{\ell -l_0}{d} -n \right)^n \Theta \left( \frac{\ell -l_0}{d} -n \right)
\label{function_f}
\end{equation}
defined for an arbitrary length $l_0$. By combining equations (\ref{V}), (\ref{Zn}), (\ref{muuu}) and (\ref{omega}), we then express the potential $V$ as
\begin{eqnarray}
\nonumber
e^{-V ( \ell ) \, d^2 /T} = 
e^{- \zeta \ell /d} + (e^{U/T} -1)^2 \; {\rm g} \left( \frac{\ell}{d} , \frac{2 l_r}{d} , \zeta \right) \\ 
-2 e^{U/T} (e^{U/T} -1) \; {\rm g} \left( \frac{\ell}{d} , \frac{l_r}{d} , \zeta \right) +e^{2U/T} \; {\rm g} \left( \frac{\ell}{d} , 0, \zeta \right) .
\label{Vinf1}
\end{eqnarray}
Using the saddle-point approximation and Stirling's formula (\ref{Stirling}), one can prove that
\begin{equation}
\lim_{\ell \to \infty} {\rm g} \left( \frac{\ell}{d} , \frac{l_0}{d} , \zeta \right) = \frac{e^{- \zeta \, l_0 /d}}{1+ \zeta} ,
\label{limit_ToProve}
\end{equation}
see \ref{appendix_proof}. We now apply this result to equation (\ref{Vinf1}) and arrive at
\begin{equation}
\lim_{\ell \to \infty} \exp \left( -V \frac{d^2}{T} \right) = \left( e^{U/T} - (e^{U/T} -1) e^{- \zeta l_r /d} \right)^2 \frac{1}{1+ \zeta} .
\end{equation}
Hence, the asymptotic value of the potential $V( \ell )$ is given by the following exact expression:
\begin{equation}
V_{\infty} = - \frac{T}{d^2} \ln \left( 1- \phi \right) -2 \frac{T}{d^2} \ln \left[ e^{U/T} - ( e^{U/T} -1) \exp \left( {- \frac{l_r}{d} \frac{\phi}{1- \phi}} \right) \right] .
\label{Vinfty}
\end{equation}
For non-adhesive particles with $U=0$ or $l_r =0$, the asymptotic value of the potential $V( \ell )$ is $V_{\infty} = -(T/d^2) \ln (1- \phi )$. For small bulk volume fractions $\phi \ll 1$ of the particles and large binding energy $U$ with $e^{U/T} \gg 1$, we obtain
\begin{equation}
V_{\infty} \approx -2 \frac{T}{d^2} \ln \left( 1 + \phi \, e^{U/T} \, \frac{l_r}{d} \right) .
\label{Vinfty2}
\end{equation}
We will use equations (\ref{Vinfty}) and (\ref{Vinfty2}) in section \ref{section_Adhesion} to calculate the effective adhesion energy of surfaces. In the following section \ref{section_transition}, we determine the global minimum of the potential $V( \ell )$. 

\section{Global minimum of the surface interaction potential \label{section_transition}}

\begin{figure}[h]
\begin{center}
\resizebox{0.7\columnwidth}{!}{\includegraphics{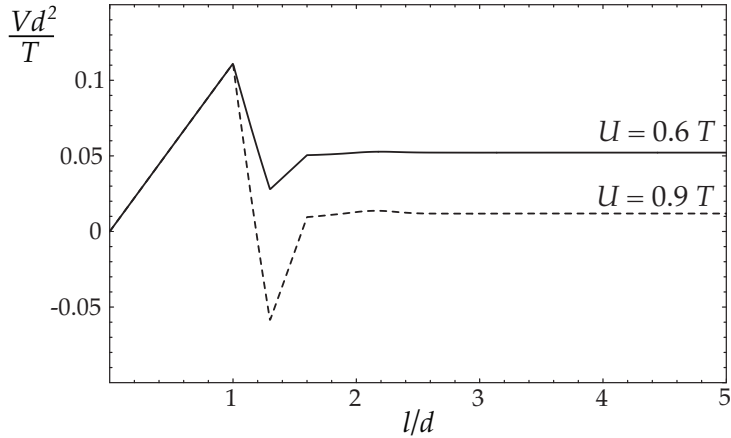}}
\caption{Rescaled surface interaction potential $V d^2/T$, given by equation (\ref{VV}), versus the rescaled surface separation $\ell /d$. 
The bulk volume fraction of the particles here is $\phi =0.1$, the binding range is $l_r =0.3 \, d$, and the binding energy is $U= 0.9 \, T$ (dashed line) and $U=0.6 \, T$ (solid line). The global minimum of the potential $V( \ell )$ is located at the separation $\ell = d+ l_r$ for $U=0.9 \, T$ (dashed line) and at surface contact $\ell =0$ for $U=0.6 \, T$ (solid line).  }
\label{V2}
\end{center}
\end{figure}

In the present calculation we have chosen the free energy reference state in such a way that the effective surface interaction potential vanishes at surface contact $\ell =0$, i.e. $V( \ell =0 )=0$, see equation~(\ref{sfe}).
For surface separations $0 < \ell <d$, the potential $V( \ell )$ increases linearly with $\ell$, i.e.~$V( \ell )= \ell T \phi /[(1- \phi) d^3]$, since $F_{\rm gc} =0$ for these separations. The potential $V( \ell )$ decreases for separations $d< \ell <d+ l_r$, and increases again for $d+ l_r < \ell < 2d$.  The potential thus attains a minimum at $\ell =d+ l_r$, see figures~\ref{V1} and \ref{V2}. The value $V_{\rm min} = V( \ell = l_r +d)$ at this minimum can be calculated again from equation~(\ref{VV}). For $\ell = d+ l_r$, equation (\ref{Zn}) reduces to $\mathcal{Z}_1 = e^{2U/T} l_r / \Lambda$ and $\mathcal{Z}_n =0$ for $n \ge 2$ since only a single particle fits into the column. Insertion of these results into equation (\ref{VV}) leads to
\begin{equation}
V_{\rm min} = - \frac{T}{d^2} \ln \left[ 1+ \frac{l_r}{d} \frac{\phi}{1- \phi} \exp \left( \frac{2 U}{T} + \frac{\phi}{1- \phi} \right) \right] + \frac{T}{d^2} \left( \frac{l_r}{d} +1 \right) \frac{\phi}{1- \phi} .
\label{V_d_plus_r}
\end{equation}

The potential $V( \ell )$ thus has two local minima, one located at $\ell =0$ with $V( \ell =0) =0$ and the other at $\ell = l_r +d$ with $V( \ell = l_r +d) = V_{\rm min}$. The two minima result from the interplay of depletion interactions and adhesive interactions. The global minimum of $V( \ell )$ is located at $\ell =d+ l_r$ if $V_{\rm min} <0$, i.e.~for
\begin{equation}
e^{- \zeta}+ \frac{l_r}{d} \, \zeta \, e^{2U/T}  - e^{  \zeta \, l_r /d } >0 
\label{inequality}
\end{equation}
with $\zeta = \phi / (1- \phi )$, see equation~(\ref{zeta}). The inequality (\ref{inequality}) is fulfilled for sufficiently large particle binding energies $U$. For small binding energies $U$, in contrast, the potential $V( \ell )$ has its global minimum at surface contact  $\ell =0$, see figure~\ref{V2}. 

In the experimentally relevant case of small particle bulk volume fractions $\phi \ll 1$ and large binding energy $U$ with $e^{U/T} \gg 1$, equation (\ref{V_d_plus_r}) reduces to 
\begin{equation}
V_{\rm min} \approx - \frac{T}{d^2} \ln \left( 1+ \phi \, e^{2U/T} \, \frac{l_r}{d} \right)
\label{Vmin2}
\end{equation}
and the inequality (\ref{inequality}) simplifies to
\begin{equation}
U  > \frac{T}{2} \ln \left( 1+ \frac{d}{l_r} \right) .
\end{equation}
and, thus, to a relation that is independent of the particle bulk volume fraction $\phi$.

\section{Adhesion energy \label{section_Adhesion}}

\begin{figure}[h]
\begin{center}
\resizebox{0.7\columnwidth}{!}{\includegraphics{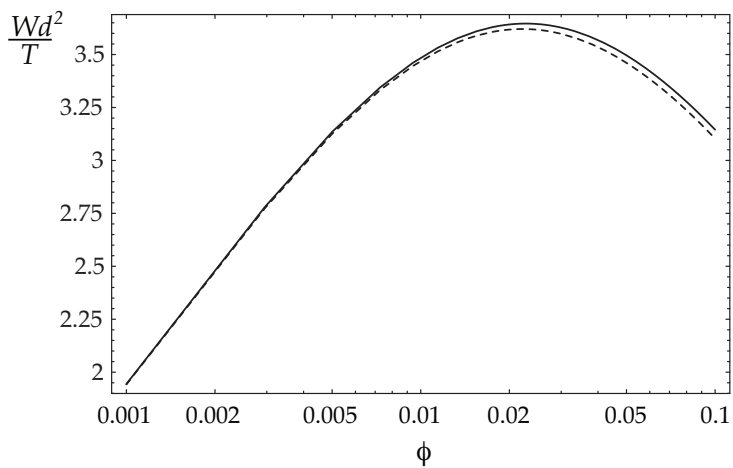}}
\caption{Rescaled adhesion energy $W d^2/T$ as a function of the particle bulk volume fraction $\phi$. The solid line corresponds to the exact result (\ref{gamma3}), and the dashed line to the approximation (\ref{gamma2}). The particle binding energy here is $U=5 \,T$ and the binding range is $l_r =0.3 \, d$. The adhesion energy has a maximum at $\phi = \phi^{\star}$ with $\phi^{\star} \approx e^{-U/T} d/l_r$. }
\label{g}
\end{center}
\end{figure}

In this section, we assume that the binding energy $U$ of the particles is sufficiently large so that the inequality (\ref{inequality}) is fulfilled. The global minimum of the interaction potential $V(\ell)$ then is located at $\ell =d+ l_r$. The minimum value $V_{\rm min} = V(\ell =d+ l_r)$ is given by equation (\ref{V_d_plus_r}). For large surface separations $\ell \gg d$, the potential $V( \ell )$ attains a constant value $V_{\infty}$ given by equation (\ref{Vinfty}). The difference between the asymptotic and the minimum value of the potential $V( \ell )$ is the effective adhesion energy
\begin{equation}
W = V_{\infty} - V_{\rm min}
\label{gamma}
\end{equation}
of the surfaces. The effective adhesion energy is the minimal work that has to performed to bring the two surfaces far apart from the separation $\ell =d+ l_r$.  From equations (\ref{Vinfty}), (\ref{V_d_plus_r}), and (\ref{gamma}), we obtain the exact result 
\begin{eqnarray}
\nonumber
W = \frac{T}{d^2} \ln
\left[ 1+ \frac{l_r}{d} \frac{\phi}{1- \phi} \exp \left( \frac{2U}{T} + \frac{\phi}{1- \phi} \right) \right]  - \frac{T}{d^2} \left( \frac{l_r}{d} +1 \right) \frac{\phi}{1- \phi} \\
- \frac{T}{d^2} \ln \left( 1- \phi \right)
- 2 \frac{T}{d^2} \ln
\left[ e^{U/T} - (e^{U/T} -1) \exp \left( - \frac{l_r}{d} \frac{\phi}{1- \phi} \right) \right] .
\label{gamma3}
\end{eqnarray}
In figure~\ref{g}, the adhesion energy $W$ is plotted as a function of the particle bulk volume fraction $\phi$. Interestingly, the adhesion energy $W$  exhibits a maximum at an optimal bulk volume fraction $\phi^{\star}$ of the particles.

\begin{figure}[h]
\begin{center}
\resizebox{0.7\columnwidth}{!}{\includegraphics{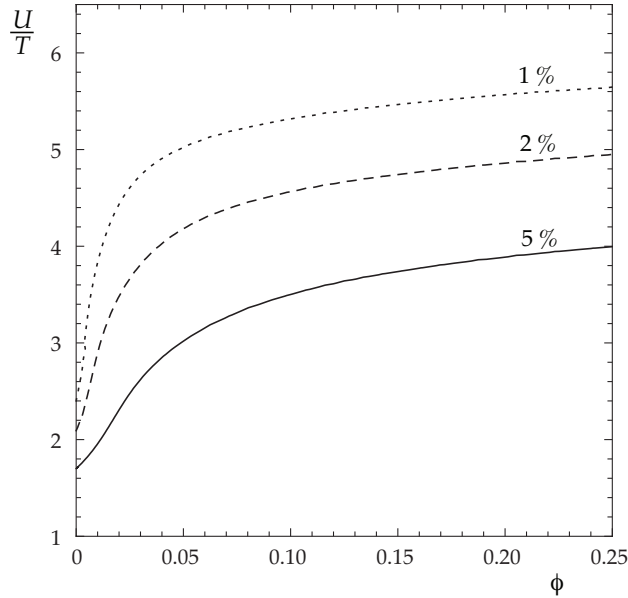}}
\caption{Relative error of the approximate expression (\ref{gamma2}) for the binding range $l_r = 0.3 /, d$ of the particles. In the parameter region above the dotted line, the relative error is smaller than 1\%. Below this line, the relative error is larger than 1\%. The relative error is smaller than 2\% above the dashed line, and smaller than 5\% above the solid line. For binding energies $U > 6 \, T$, the simple expression (\ref{gamma2}) approximates the exact result (\ref{gamma3}) very well since the relative error is smaller than 1\% for a broad range of volume fractions $\phi$.}
\label{W0W}
\end{center}
\end{figure}

For small bulk volume fractions $\phi\ll 1$ and large binding energy $U$ with $e^{U/T} \gg 1$, the asymptotic value and minimum value of $V(\ell)$ are approximately given by equations (\ref{Vinfty2}) and (\ref{Vmin2}), respectively. The adhesion energy $W= V_{\infty} - V_{\rm min}$ then simplifies to
\begin{equation}
W \approx \frac{T}{d^2} \ln \frac{ 1+ \phi \, e^{2U/T} \, l_r /d }{ \left( 1+ \phi \, e^{U/T} \, l_r /d \right)^2 } .
\label{gamma2}
\end{equation}
This expression is identical with our previous result obtained from a virial expansion in $\phi$ up to second order terms \cite{Rozycki08}. For $\phi \ll 1$ and $e^{U/T} \gg 1$, the adhesion energy (\ref{gamma2}) is a good approximation of the exact result (\ref{gamma3}), see figures~\ref{g} and \ref{W0W}. From equation~(\ref{gamma2}), we obtain the approximate expression
\begin{equation}
\phi^{\star} \approx \frac{d}{l_r} \, e^{-U/T} 
\label{X_b^star}
\end{equation}
for the optimum bulk volume fraction $\phi^{\star}$ at which the adhesion energy $W$ attains its maximum value. 

The adhesion energy (\ref{gamma2}) can be understood as the difference of two Langmuir adsorption free energies per fluid column, or pair of apposing binding sites \cite{Rozycki08}: (i) the adsorption free energy $(T/d^2)\ln \left( 1+ q\, \phi \, e^{2U/T} \right)$ for small surface separations at which a particle binds both surfaces with total binding energy $2U$, and (ii) the adsorption free energy $(T/d^2)\ln \left(1+ q \, \phi \, e^{U/T} \right)$ for large surface separations, counted twice in (\ref{gamma2}) because we have two surfaces. These Langmuir adsorption free energies  result from a simple two-state model in which a particle is either absent (Boltzmann weight $1$) or present (Boltzmann weights $q\, \phi \, e^{2U/T}$ and $q\, \phi \, e^{U/T}$, respectively) at a given binding site, see e.g.~\cite{davis96}. The factor $q$ depends on the degrees of freedom of a single adsorbed particle. In our model, we obtain $q=l_r/ d$.

To assess the quality of approximate expression (\ref{gamma2}), we analyze its relative error in reference to the exact result (\ref{gamma3}). The relative error is the magnitude of the difference between the exact result (\ref{gamma3}) and the approximate expression (\ref{gamma2}) divided by the magnitude of the exact result (\ref{gamma3}). Figure~\ref{W0W} shows parameter regions in which the relative error of the expression (\ref{gamma2}) is smaller or larger than 1\%, 2\% and 5\%, respectively. In this example, the binding range is $l_r = 0.3$. For intermediate and large binding energies with $U > 6 \, T$, we find that the relative error of the approximate expression (\ref{gamma2}) is smaller than 1\% in a broad range of volume fractions~$\phi$. 

\section{Potential barrier \label{section_barrier}}

For large binding energies $U$ with $e^{U/T} \gg 1$, the effective interaction potential has a barrier at surface separations $d+2 l_r< \ell < 2(d+ l_r )$, see figure~\ref{V1}. At these separations, only a single particle fits between the surfaces, but this particle can just bind one of the surfaces. The particle thus 'blocks' the binding site at the apposing surface. 

The potential barrier attains its maximum value $V_{\rm ba}=  V( \ell = 2d)$ at the separation $\ell =2d$, see figure~\ref{V1}. From equation (\ref{Vinf1}), we obtain
\begin{equation}
V_{\rm ba} = - \frac{T}{d^2} \ln \left[ 1+ \left( 2 \frac{l_r}{d} \left( e^{U/T} -1 \right) +1 \right) \frac{\phi}{1- \phi} \exp \left( \frac{\phi}{1- \phi} \right) \right] +2 \frac{T}{d^2} \frac{\phi}{1- \phi}
\end{equation}
For $\phi\ll 1$ and $e^{U/T} \gg 1$, we get
\begin{equation}
V_{\rm ba} \approx - \frac{T}{d^2} \ln \left( 1+ 2 \phi \, e^{U/T} \, \frac{l_r}{d}  \right) 
\end{equation}
The barrier height $U_{\rm ba} = V_{\rm ba} - V_{\infty} $ then is  
\begin{equation}
U_{\rm ba} \approx \frac{T}{d^2} \ln \frac{\left( 1+ \phi \, e^{U/T} \, l_r/d \right)^2}{1+ 2 \phi \, e^{U/T} \, l_r/d } 
\label{U_ba}
\end{equation}
since the asymptotic value $V_{\infty}$ is given by equation (\ref{Vinfty2}) in this limiting case. The width of the barrier is approximately $l_{\rm ba} \approx d$, see figure~\ref{V1}. Equation (\ref{U_ba}) is again identical with our previous result  obtained from a virial expansion in $\phi$ up to second order terms \cite{Rozycki08}.

\section{Binding probability}

Another quantity of interest here is the binding probability
\begin{equation}
n_s = \frac{1}{2} \langle \Theta \left( l_r - x_1 + d/2 \right) \rangle + \frac{1}{2} \langle \Theta \left( x_n - \ell +l_r + d/2 \right) \rangle
\label{cc}
\end{equation}
defined as the probability that the separation of the closest particle from a column base is smaller than the binding range $l_r$. In other words, the binding probability $n_s$ is the probability of finding a particle bound to one of the bases. The binding probability corresponds to the surface coverage in the case of a three-dimensional gas of particles between two parallel attractive surfaces. 

Equations~(\ref{square-well-pot}) - (\ref{fe}) imply that the binding probability can be calculated by differentiation of the grand potential $F_{\rm gc}$ with respect to binding energy $U$, i.e. $n_s = - \frac{1}{2} \left( \frac{\partial F_{\rm gc}}{\partial U} \right) $. Since the grand potential density $f_{\rm gc}$ given by equation (\ref{omega}) does not depend on the binding energy $U$, the binding probability can also be obtained from the effective surface interaction potential {\em via}
\begin{equation}
n_s = - \frac{d^2}{2} \left( \frac{\partial V}{\partial U} \right) .
\label{cov_def}
\end{equation}

\begin{figure}[h]
\begin{center}
\resizebox{0.7\columnwidth}{!}{\includegraphics{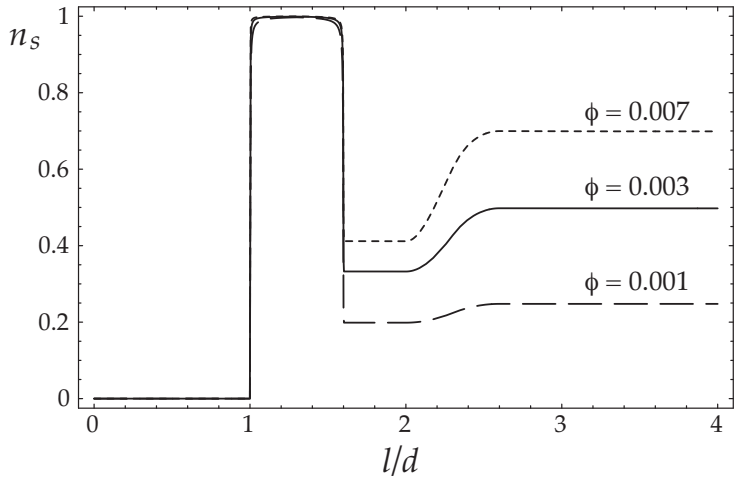}}
\caption{Binding probability $n_s$, calculated numerically from equations (\ref{cov_def}) and (\ref{Vinf1}), as a function of rescaled surface separation $\ell /d$ where $d$ is the diameter of the adhesive particles. The binding energy here is $U=7 \, T$, the binding range is $l_r = 0.3 \, d$ and the particle bulk volume fraction is $\phi =0.001 < \phi^{\star}$ (dashed line), $\phi =0.003 \approx \phi^{\star}$ (solid line) and $\phi =0.007 > \phi^{\star}$ (dotted line). The optimal volume fraction $\phi^{\star}$ at which the adhesion energy becomes maximal is given by equation (\ref{X_b^star}). }
\label{c}
\end{center}
\end{figure}

With the exact expression (\ref{Vinf1}) for the interaction potential $V$, the binding probability $n_s$ can be determined numerically for any finite separation $\ell$. In figure~\ref{c}, the binding probability $n_s$ is plotted as a function of surface separation $\ell$ for three different volume fractions $\phi$ around the optimal volume fraction $\phi^{\star}$ at which the adhesion energy $W$ is maximal. In the vicinity of $\phi^{\star}$, the binding probability at large separations $\ell > 2(d+l_r)$ is sensitive to small variations of $\phi$, while the binding probability at separations $\ell$ in the surface binding range $d< \ell < d+ 2l_r$ remains practically constant at almost 100\% .

For small bulk volume fractions $\phi$ of the particles and large particle binding energies $U$, the asymptotic and minimum value of the interaction potential $V( \ell )$ are given by equations (\ref{Vinfty2}) and (\ref{Vmin2}), respectively. From these equations and relation (\ref{cov_def}), we obtain the approximate expressions
\begin{equation}
n_{s, \, \infty} \approx \frac{\phi}{\phi + \phi^{\star} }
\label{cov_infty}
\end{equation}
for the binding probability at large surface separation $\ell$ and
\begin{equation}
n_{s, \, \rm min} \approx \frac{\phi}{\phi + \phi^{\star} e^{-U/T} } ,
\label{cov_min}
\end{equation}
for the particle binding probability at the binding separation $\ell =d+ l_r$ of the surfaces, with the optimum bulk volume fraction $\phi^{\star}$ given in equation (\ref{X_b^star}). These expressions correspond to the well-known Langmuir adsorption equation \cite{davis96}. At the optimal volume fraction $\phi^{\star}$, the particle binding probability for unbound and bound surfaces is $n_{s, \, \infty} = 1/2$ and $n_{s, \, \rm min} \approx 1$, respectively. Bringing the surfaces from large separations $\ell > 2(d+ l_r)$ within binding separations $d< \ell < d+ 2l_r$ thus does not require desorption or adsorption of particles at $\phi=\phi^{\star}$.

\section{Conclusions}

We have considered one-dimensional gas of hard-sphere particles with attractive boundaries, a novel extension of the Tonks model \cite{tonks36}. We have solved this model analytically in the whole range of parameters by explicit integration over the particles' degrees of freedom in the partition function. In contrast to other studies on one-dimensional models for hard spheres \cite{salsburg53, baur62, davis96, davis90, henderson07, lekkerkerker00, wensink04, vanderSchoot96, chou03}, we have focused on the boundary contribution to the free energy of the system, which corresponds to the effective, particle-mediated interaction potential between the boundaries, or surfaces, see figures~\ref{V1} and~\ref{V2}. The effective adhesion energy obtained from the interaction potential depends non-monotonically on the volume fraction $\phi$ of the particles in the bulk, see figure~\ref{g}. The adhesion energy exhibits a maximum at an optimum volume fraction, which can lead to reentrant transitions in which the surfaces first bind with increasing volume fraction $\phi$, and unbind again when the volume fraction $\phi$ is increased beyond its optimum value. 

A lattice of such one-dimensional gas columns represents a discrete approximation of a three-dimensional gas of particles between two adsorbing surfaces, see figure~\ref{cartoon} and reference~\cite{Rozycki08}. For small volume fractions $\phi$ and short-ranged particle-surface interactions considered here, the gas of particles between two well-separated surfaces is as dilute as in the bulk, except for the single adsorption layers of particles at the surfaces. At larger volume fractions, three-dimensional packing effects become relevant. These effects are not captured correctly in the one-dimensional model. However, it has been pointed out \cite{henderson07} that approximations based on one-dimensional models do well in comparison to density functional theories for three-dimensional hard sphere fluids confined in planar, non-adsorbing pores \cite{pizio01}. In principle, the quality of the one-dimensional approximation can be tested by Monte Carlo or Molecular Dynamics simulations, which have been used to study various three-dimensional systems of hard spheres confined between non-adsorbing surfaces \cite{chu94, schmidt97, schoen98, fortini06, mittal08}. 

For simplicity, and for consistency with our previous publication \cite{Rozycki08}, we have considered here a square lattice of columns between the surfaces. In particular, the factor $d^2$ in the denominator of equation~(\ref{sfe}) is the column base area in the square lattice. For a hexagonal lattice of columns, the corresponding area is $( \sqrt{3} /2 ) \, d^2$, and the corresponding effective interaction potential of the surfaces is thus obtained by multiplying the right hand side of equation~(\ref{VV}) with a factor $2 / \sqrt{3} \approx 1.1547$. The adhesion energy of the surfaces has to be rescaled with the same factor in the case of hexagonal lattice of columns, but its functional dependence on the bulk volume fraction $\phi$, binding energy $U$, and binding range $l_r$ remains unchanged.

We have considered an equilibrium situation in which the particles exchange with a bulk solution. For polymers between surfaces, such an equilibrium has been termed `full equilibrium'. In a `restricted equilibrium', in contrast, the polymers are trapped between the surfaces \cite{ennis99,bleha04,leermakers05}, which is less likely for the spherical particles considered here.

\appendix

\section{Canonical partition function \label{appendix_cpf} }

In this section we calculate the $n$-particle partition function $\mathcal{Z}_n$ as given by equation~(\ref{integral_Zn}) for $\ell > nd$. First, if one notices that
\begin{equation}
e^{U \, \Theta ( l_r - y_1 ) /T} = 1+ \left( e^{U/T} -1 \right) \Theta ( l_r - y_1 )
\end{equation}
and
\begin{equation}
e^{U \, \Theta ( y_n - \ell + nd + l_r ) /T} = 1+ \left( e^{U/T} -1 \right) \Theta ( y_n - \ell + nd + l_r ) ,
\end{equation}
the integral (\ref{integral_Zn}) can be written as a sum of four terms
\begin{equation}
\mathcal{Z}_n = \frac{1}{\Lambda^n} I_1 + \frac{1}{\Lambda^n} \left( e^{U/T} -1 \right) I_2 + \frac{1}{\Lambda^n} \left( e^{U/T} -1 \right) I_3 + \frac{1}{\Lambda^n} \left( e^{U/T} -1 \right)^2 I_4 ,
\label{I_split}
\end{equation}
with
\begin{equation}
I_1 =
\int_{0}^{\ell -nd} {\rm d} y_n \int_{0}^{y_n} {\rm d}y_{n-1} \dots \int_{0}^{y_3} {\rm d}y_2 \int_{0}^{y_2} {\rm d} y_1 ,
\end{equation}
\begin{equation}
I_2 = \int_{0}^{\ell -nd} {\rm d} y_n \Theta \left( y_n - \ell + nd + l_r \right) \int_{0}^{y_n} {\rm d}y_{n-1} \dots \int_{0}^{y_3} {\rm d}y_2 \int_{0}^{y_2} {\rm d} y_1 ,
\end{equation}
\begin{equation}
I_3 = \int_{0}^{\ell -nd} {\rm d} y_n \int_{0}^{y_n} {\rm d}y_{n-1} \dots \int_{0}^{y_3} {\rm d}y_2 \int_{0}^{y_2} {\rm d} y_1 \Theta \left( l_r - y_1 \right) ,
\end{equation}
\begin{equation}
I_4 = \int_{0}^{\ell -nd} {\rm d} y_n \Theta \left( y_n - \ell + nd + l_r \right) \int_{0}^{y_n} {\rm d}y_{n-1} \dots \int_{0}^{y_2} {\rm d} y_1 \Theta \left( l_r - y_1 \right) .
\end{equation}
The first integral 
\begin{equation}
I_1 = \frac{1}{(n-1)!} \int_{0}^{\ell -nd} y_n^{n-1} {\rm d} y_n = \frac{1}{n!} \left( \ell -nd \right)^n
\label{I1a}
\end{equation}
and second integral
\begin{eqnarray}
\nonumber
I_2 = \frac{1}{(n-1)!} \int_{0}^{\ell -nd} y_n^{n-1} \Theta \left( y_n - \ell + nd + l_r \right) {\rm d} y_n \\ 
= \frac{1}{n!} \left( \left( \ell - nd \right)^n - \left( \ell - nd - l_r \right)^n \Theta \left( \ell - nd - l_r \right) \right)
\label{I2a}
\end{eqnarray}
can be easily calculated.

To calculate the third integral, we start from
\begin{equation}
\int_0^{y_2} \Theta \left( l_r - y_1 \right) {\rm d} y_1 = \min \left[ y_2, l_r \right] = y_2 - \left( y_2 - l_r \right) \Theta \left( y_2 - l_r \right) .
\label{help_int_1}
\end{equation}
In the next steps, we find
\begin{equation}
\int_0^{y_3} \left( y_2 - \left( y_2 - l_r \right) \Theta \left( y_2 - l_r \right) \right) {\rm d} y_2 = \frac{1}{2} y_3^2 - \frac{1}{2} \left( y_3 - l_r \right)^2 \Theta \left( y_3 - l_r \right)
\label{help_int_2}
\end{equation}
and
\begin{equation}
\int_0^{y_4} \left( \frac{1}{2} y_3^2 - \frac{1}{2} \left( y_3 - l_r \right)^2 \Theta \left( y_3 - l_r \right) \right) {\rm d} y_3 = \frac{1}{6} y_4^3 - \frac{1}{6} \left( y_4 - l_r \right)^3 \Theta \left( y_4 - l_r \right)
\label{help_int_3}
\end{equation}
Iterating these results leads to
\begin{equation}
I_3 = \frac{1}{(n-1)!} \int_0^{\ell -nd} \left[ y_n^{n-1} - \left( y_n - l_r \right)^{n-1} \Theta \left( y_n - l_r \right) \right] {\rm d} y_n .
\end{equation}
The integral $I_3$ can now be evaluated as
\begin{eqnarray}
\nonumber
I_3 = \frac{1}{n!} \left( \ell - nd \right)^n - \frac{1}{(n-1)!} \int_0^{\ell -nd} \left( y_n - l_r \right)^{n-1} \Theta \left( y_n - l_r \right) {\rm d} y_n \\
= \frac{1}{n!} \left( \ell - nd \right)^n - \frac{1}{n!} \left( \ell - nd - l_r \right)^n \Theta \left( \ell - nd - l_r \right) .
\label{I3a}
\end{eqnarray}
Note that $I_2 = I_3$. 

The fourth integral, $I_4$, can be brought to the form
\begin{equation}
I_4 = \frac{1}{(n-1)!} \int_0^{\ell -nd} \Theta \left( y_n - \ell +nd + l_r \right) \left[ y_n^{n-1} - \left( y_n - l_r \right)^{n-1} \Theta \left( y_n - l_r \right) \right] {\rm d} y_n
\label{I4a}
\end{equation}
if one uses again (\ref{help_int_1}) and iterates the integration as in (\ref{help_int_2}) and (\ref{help_int_3}). The first term on the right hand side of equation (\ref{I4a}) is equal to $I_2$, see equation~(\ref{I2a}). Thus 
$ I_4 = I_2 - I_5 $, where
\begin{equation}
I_5 =  \frac{1}{(n-1)!} \int_0^{\ell -nd} \Theta \left( y_n - \ell +nd + l_r \right) \Theta \left( y_n - l_r \right) \left( y_n - l_r \right)^{n-1} {\rm d} y_n
\end{equation}
To determine the integral $I_5$, one has to distinguish three cases: (i) for $\ell > nd +2 l_r$, we obtain
\begin{equation}
I_5 = \frac{1}{n!} \left( \left( \ell -nd - l_r \right)^n - \left( \ell -nd -2 l_r \right)^n \right) ,
\end{equation}
(ii) for $nd+ l_r < \ell < nd +2 l_r$, we obtain
\begin{equation}
I_5 = \frac{1}{n!} \left( \ell -nd - l_r \right)^n
\end{equation}
and (iii) for $\ell < nd + l_r$ one gets $I_5 =0$. In summary
\begin{eqnarray}
\nonumber
I_4 = I_2 - \frac{1}{n!} \left( \ell -nd - l_r \right)^n \Theta \left( \ell - nd - l_r \right) \\
+ \frac{1}{n!} \left( \ell -nd -2 l_r \right)^n \Theta \left( \ell - nd -2 l_r \right)
\label{I4b}
\end{eqnarray}
If one now gathers the results (\ref{I1a}), (\ref{I2a}), (\ref{I3a}), (\ref{I4b}) and returns to equation (\ref{I_split}), one obtains the partition function
\begin{equation}
\mathcal{Z}_n = \frac{1}{\Lambda^n n!} \left[ J_1 + 2( e^{U/T} -1 )( J_1 - J_2 ) +( e^{U/T} -1 )^2 ( J_1 -2 J_2 + J_3 ) \right]
\end{equation}
with
\begin{equation}
J_1= \left( \ell -nd \right)^n  \Theta \left( \ell -nd \right) ,
\end{equation}
\begin{equation}
J_2= \left( \ell -nd - l_r \right)^n  \Theta \left( \ell -nd - l_r \right) ,
\end{equation}
\begin{equation}
J_3= \left( \ell -nd -2 l_r \right)^n  \Theta \left( \ell -nd -2 l_r \right) .
\end{equation}
This result can be written as
\begin{eqnarray}
\nonumber
\mathcal{Z}_n =
\frac{1}{\Lambda^n n!} \bigg[
\left(e^{U/T} -1 \right)^2 \left( \ell - nd -2 l_r \right)^n \Theta \left( \ell - nd -2 l_r \right)  \\
\nonumber
- 2 e^{U/T} \left( e^{U/T} -1 \right) \left( \ell - nd - l_r \right)^n \Theta \left( \ell - nd - l_r \right) \\
+ e^{2U/T} \left( \ell - nd \right)^n \Theta \left( \ell - nd \right) \bigg]
\end{eqnarray}
which simplifies to equation (\ref{Zn}).

\section{ Proof of equality (\ref{limit_ToProve}) \label{appendix_proof} }

Here, we explore the asymptotics of the function g$( \ell /d, l_0 /d, \zeta)$ defined in equation (\ref{function_f}) and, hence, prove the equality (\ref{limit_ToProve}). To simplify the notation, let $\ell /d = N$, where $N$ is a large integer number, and $l_0 /d = \lambda$. Then
\begin{equation}
{\rm g} \left( N, \lambda, \zeta \right)= \sum_{n=1}^{N} \frac{\zeta^n}{n!} e^{ -(N-n) \zeta} \left( N -n -\lambda \right)^n \Theta \left( N -n - \lambda \right) .
\label{function_ff}
\end{equation}
In the next step, we introduce the auxiliary function
\begin{equation}
\chi (n) = -(N-n) \zeta + n \ln \zeta + n \ln (N -n - \lambda) - n \ln n + n - \frac{1}{2} \ln ( 2 \pi n )
\label{function_phi}
\end{equation}
and rewrite the function g$(N, \lambda, \zeta)$ given by equation (\ref{function_ff}) in the form
\begin{equation}
{\rm g} = \sum_{n=1}^{N} e^{ \chi (n) } \; \Theta \left( N -n - \lambda \right) ,
\label{function_fff}
\end{equation}
using Stirling's formula (\ref{Stirling}). For large $N$, one can replace the sum on the right hand side of equation (\ref{function_fff}) by an integral and write
\begin{equation}
{\rm g} \approx \int_{1}^{N- \lambda} e^{ \chi (n) } \, {\rm d} n .
\end{equation}
The function $\chi (n)$ has a global maximum at $n=n_0$ with
\begin{equation}
n_0 \approx \left( N- \lambda \right) \frac{\zeta}{1+ \zeta }
\label{n0}
\end{equation}
for large $N$. Note that for large numbers $N$, the location $n_0 \approx \left( N- \lambda \right) \phi$ of the global minimum scales linearly with $N$. If we now expand the function $\chi (n)$ around the point $n_0$ up to second order terms and apply the saddle-point approximation, we get
\begin{equation}
{\rm g} \approx e^{\chi ( n_0 )} \int_{1}^{N- \lambda} \exp \left[ \frac{1}{2} \chi^{''} ( n_0 ) \, \left( n- n_0 \right)^2 \right] {\rm d} n .
\end{equation}
A simple change of variables $m=n-n_0$ leads to
\begin{equation}
{\rm g} \approx e^{\chi ( n_0 )} \int_{-(N- \lambda ) \phi +1 }^{ (N- \lambda )(1- \phi ) } e^{ \chi^{''} ( n_0 ) \, m^2 /2} \; {\rm d} m
\end{equation}
where we have used $n_0 \approx \left( N- \lambda \right) \phi$, which follows from equation (\ref{n0}) and relation (\ref{zeta}) between variables $\zeta$ and $\phi$. In the limit of large $N$, we thus obtain
\begin{equation}
{\rm g} \approx e^{\chi ( n_0 )} \int_{- \infty}^{\infty} e^{ \chi^{''} ( n_0 ) \, m^2 /2} \; {\rm d} m
\label{function_f_sp}
\end{equation}
because $0< \phi <1$. Now, we can calculate the Gaussian integral in (\ref{function_f_sp}) to get
\begin{equation}
{\rm g} \approx e^{\chi ( n_0 )} \sqrt{ \frac{2 \pi}{ - \, \chi^{''} (n_0) } \; } .
\label{function_f_G}
\end{equation}
From the definition (\ref{function_phi}) of the auxiliary function $\chi (n)$ and equation (\ref{n0}) for the point $n=n_0$ at which function $\chi (n)$ has its global maximum, we get
\begin{equation}
\chi ( n_0 ) \approx - \lambda \, \zeta - \frac{1}{2} \ln ( 2 \pi n_0 )
\label{phi_n0}
\end{equation}
and
\begin{equation}
\chi^{''} ( n_0 ) \approx - \frac{ \left( 1+ \zeta \right)^2 }{n_0 }
\label{phi_bis_n0}
\end{equation}
Note that $\chi^{''} ( n_0 ) <0$, and that the function $\chi (n)$ has indeed a maximum at $n=n_0$. Combining equations (\ref{function_f_G}), (\ref{phi_n0}) and (\ref{phi_bis_n0}) leads to
\begin{equation}
{\rm g} \approx \frac{ e^{- \lambda \, \zeta} }{1+ \zeta }
\end{equation}
for large $N$ values and, thus, to equation (\ref{limit_ToProve}) {\sl quod erat demonstrandum}.

\end{document}